\documentclass[preprint,preprintnumbers,amsmath,amssymb]{revtex4}
\pdfoutput=1
\usepackage{dcolumn}
\usepackage{bm}
\usepackage{graphicx}
\usepackage{booktabs}
\usepackage{makecell}
\usepackage{array}

\begin{document}

\title{Strong cosmic censorship for the massless Dirac field in the Reissner-Nordstrom-de Sitter spacetime}
\author{Boxuan Ge$^{1,2}$}
\thanks{boxuange@mail.bnu.edu.cn}
\author{Jie Jiang$^2$}
\thanks{jiejiang@mail.bnu.edu.cn}
 \author{Bin Wang$^{1,3}$}
 \thanks{wangb@yzu.edu.cn}
\author{Hongbao Zhang$^{2,4}$}
\thanks{hzhang@vub.ac.be}
 \author{Zhen Zhong$^{2}$}
 \thanks{zhenzhong@mail.bnu.edu.cn}

\affiliation{
 $^1$ Center for Gravitation and Cosmology, College of Physical Science and Technology,
Yangzhou University, Yangzhou 225009, China\\
$^2$ Department of Physics, Beijing Normal University, Beijing 100875, China\\
$^3$ Department of Physics and Astronomy, Shanghai Jiao Tong University,
Shanghai 200240, China\\
 $^4$ Theoretische Natuurkunde, Vrije Universiteit Brussel,
 and The International Solvay Institutes, Pleinlaan 2, B-1050 Brussels, Belgium
 }

\begin{abstract}
We present the Fermi story of strong cosmic censorship in the near-extremal Reissner-Nordstrom-de Sitter black hole. To this end, we first derive from scratch the criterion for the quasi-normal modes of Dirac field to violate strong cosmic censorship in such a background, which turns out to be exactly the same as those for Bose fields, although the involved energy momentum tensor is qualitatively different from that for Bose fields.  Then to extract the low-lying quasi-normal modes by Prony method, we apply Crank-Nicolson method to evolve our Dirac field in the double null coordinates. As a result, it shows that for a fixed near-extremal black hole,  strong cosmic censorship can be recovered by the $l=\frac{1}{2}$ black hole family mode once the charge of our Dirac field is greater than some critical value, which is increased as one approaches the extremal black hole.

\end{abstract}

\maketitle

\section{Introduction and motivation}
The law of physics is called well posed if and only if the future data of the involved physical entities can be determined uniquely from the appropriately prescribed initial conditions. Otherwise, the law of physics would be void of its  predictability. However, the appearance of Cauchy horizons in some solutions to Einstein equation seems to make general relativity lose its predictability across the Cauchy horizons. To maintain the predictability of general relativity, long time ago Penrose proposed his strong cosmic censorship conjecture (SCC), which asserts that such Cauchy horizons are unstable under perturbations. As a result, the would-be Cauchy horizons become the singular boundaries of spacetime such that everything including Einstein equation can not be extended across them. It is noteworthy that there are different mathematical formulations of SCC, according to what it precisely means by saying whether the Cauchy horizon is extendible or not. Hereafter we shall focus exclusively on the Christodoulou's formulation\cite{Christodoulou}, because it is formulated to most conforms to our physical intuition.

It is fair to say that SCC remains in good health until the very recent work by Cardoso and his companions\cite{CCDHJ1}. They find that for a near-extremal Reissner-Nordstrom-de Sitter (RNdS) black hole, the linear massless neutral scalar field perturbations do not cause the Cauchy horizon generically inextendible due to the fact that the blue shift amplification along the Cauchy horizon is overshadowed by the exponential decay behavior outside of the black hole in de Sitter space. Thus SCC is violated. This has stimulated a series of works to examine the validity of SCC in de Sitter space\cite{DERS,Hod1,Hod2,Hod,CCDHJ2,DRS1,MTWZZ,DRS2,LZCCN}. To be more specific, on the one hand, it is found that totally different from the case for the RNdS black hole, no violation of SCC occurs for the linear perturbations on top of the Kerr-de Sitter black hole and sufficiently rapid rotating Kerr-Newman-de Sitter black hole\cite{DERS,Hod1}. On the other hand, it is found that the coupled linear electromagnetic and gravitational perturbations give rise to a much worse violation of SCC on top of the RNdS black hole\cite{DRS1}. Faced up with such violations on top of the RNdS black hole, one may first ask whether there are some ways to recover SCC. Note that as alluded to above, such violations come from the linear analysis. So one potential way out is to go to the full nonlinear level. As a result, it is shown that the nonlinear effects do not suffice to save SCC for the scalar perturbations\cite{LZCCN}. However, taking into account the fact that the formation of a charged black hole entails the presence of the remnant charged fields, one is required to see what happens to the charged perturbations on top of the RNdS black hole\cite{Hod2}. The relevant results show that SCC can be recovered for the charged scalar perturbations except in the highly extremal limit, where there still exists room for the violation of SCC\cite{Hod,CCDHJ2,MTWZZ,DRS2}. However, all of the above considerations restrict into the perturbations from the Bose fields. So a natural question along this line is what happens to the perturbations from the Fermi fields. In particular, the purpose of this paper is see whether the massless Dirac field perturbations can save SCC out of the RNdS black hole.

The structure of this paper is organized as follows. In the subsequent section, we derive from scratch the relationship between the quasinormal modes (QNMs) and SCC for Dirac field in RNdS black hole. In Section \ref{numerics}, after developing our numerical scheme for the time evolution of the charged Dirac field in the double null coordinates by Crank-Nicolson method, we present the relevant numerical results about the low-lying QNMs for the massless Dirac field, which demonstrates that SCC can be recovered when the charge of our Dirac field is greater than a critical value for a fixed near-extremal RNdS black hole. We conclude our paper in the last section. We relegate the WKB approximation for the large $l$ and large $q$ limit into Appendix A and B, respectively.

\section{Quasi-normal modes and strong cosmic censorship}\label{qnm}
Start with the $4$-dimensional RNdS black hole
\begin{equation}
ds^2=f(r)dt^2-\frac{dr^2}{f(r)}-r^2(d\theta^2+\sin^2\theta d\varphi^2), \quad A_a=-\frac{Q}{r}(dt)_a,
\end{equation}
where
\begin{equation}
f(r)=1-\frac{2M}{r}+\frac{Q^2}{r^2}-\frac{\Lambda r^2}{3}
\end{equation}
with $M$ and $Q$ the mass and charge of the black hole, and $\Lambda$ the positive cosmological constant. If the cosmological, event, and Cauchy horizons are denoted as $r_c$, $r_+$, and $r_-$ individually, then we can rewrite
$f(r)=\frac{\Box}{r^2}$, where $\Box=\frac{\Lambda}{3}(r_c-r)(r-r_+)(r-r_-)(r-r_o)$ with $r_o=-(r_c+r_++r_-)$. In addition, the surface gravity at each horizon $r_h$ is defined as $\kappa_h=|\frac{1}{2}f'(r_h)|$. Accordingly, we have
\begin{eqnarray}
\kappa_c&=&\frac{\Lambda}{6r_c^2}(r_c-r_+)(r_c-r_-)(r_c-r_o),\nonumber\\
\kappa_+&=&\frac{\Lambda}{6r_+^2}(r_c-r_+)(r_+-r_-)(r_+-r_o),\nonumber\\
\kappa_-&=&\frac{\Lambda}{6r_-^2}(r_c-r_-)(r_+-r_-)(r_--r_o),\nonumber\\
\kappa_o&=&\frac{\Lambda}{6r_o^2}(r_c-r_o)(r_+-r_o)(r_--r_o).
\end{eqnarray}

In such a curved spacetime, the action for the Dirac field is given by
\begin{equation}
S=i\int d^4x\sqrt{-g}\frac{1}{2}(\bar{\phi}^{A'}D_{A'A}\phi^A-\phi^A\bar{D}_{AA'}\bar{\phi}^{A'}
+\bar{\sigma}_AD^{AA'}\sigma_{A'}-\sigma_{A'}\bar{D}^{A'A}\bar{\sigma}_A)
-\frac{\mathcal{M}}{\sqrt{2}}(\bar{\phi}^{A'}\sigma_{A'}-\phi^A\bar{\sigma}_A),
\end{equation}
where $D_{AA'}=\nabla_{AA'}-iqA_{AA'}$ with $\mathcal{M}$ and $q$ the mass and charge of our Dirac field. The variation of the action gives rise to the Dirac equation as
\begin{equation}
D_{A'A}\phi^A=\frac{\mathcal{M}}{\sqrt{2}}\sigma_{A'},\quad D^{AA'}\sigma_{A'}=-\frac{\mathcal{M}}{\sqrt{2}}\phi^A,
\end{equation}
which keeps invariant not only under the gauge transformation $(A_a, \phi^A,\sigma_{A'})\rightarrow (A_a+(d\lambda)_a, e^{iq\lambda}\phi^A,e^{iq\lambda}\sigma_{A'})$ but also under the charge conjugation $(q,\phi^A,\sigma_{A'})\rightarrow (-q,\bar{\sigma}^A,\bar{\phi}_{A'})$.

In terms of the dyad $(\xi_1^A=o^A,\xi_2^A=\iota^A)$ with $o_A\iota^A=1$, the above Dirac equation can be expressed as

\begin{eqnarray}
{[D+\rho-\varepsilon-iq(A\cdot n)]}\phi^1+{[\delta+\alpha-\pi-iq(A\cdot m)]}\phi^2&=&\frac{\mathcal{M}}{\sqrt{2}}\sigma_1,\nonumber\\
{[\bar{\delta}+\tau-\beta-iq(A\cdot\bar{m})]}\phi^1+{[\Delta+\gamma-\mu-iq(A\cdot l)]}\phi^2&=&\frac{\mathcal{M}}{\sqrt{2}}\sigma_2,\nonumber\\
{-[\bar{\delta}+\bar{\alpha}-\bar{\pi}-iq(A\cdot\bar{m})]}\sigma_1+{[D+\bar{\rho}-\bar{\varepsilon}-iq(A\cdot n)]}\sigma_2&=&-\frac{\mathcal{M}}{\sqrt{2}}\phi^2,\nonumber\\
{-[\Delta+\bar{\gamma}-\bar{\mu}-iq(A\cdot l)]}\sigma_1+{[\delta+\bar{\tau}-\bar{\beta}-iq(A\cdot m)]}\sigma_2&=&\frac{\mathcal{M}}{\sqrt{2}}\phi^1,
\end{eqnarray}
where the derivative operators and spin-coefficients are defined as
\begin{equation}
D=n^a\nabla_a,\quad \Delta=l^a\nabla_a,\quad \delta=m^a\nabla_a,\quad \bar{\delta}=\bar{m}^a\nabla_a,
\end{equation}
and
\begin{eqnarray}
&&\kappa=o_ADo^A,\quad \varepsilon=o_AD\iota^A,\quad \pi=\iota_AD\iota^A,\quad \tau=o_A\Delta o^A,\quad \gamma=o_A\Delta\iota^A,\quad \nu=\iota_A\Delta\iota^A,\nonumber\\
&&\rho=o_A\delta o^A,\quad \alpha=o_A\delta\iota^A,\quad\lambda=\iota_A\delta\iota^A,\quad \sigma=o_A\bar{\delta}o^A,\quad,\beta=o_A\bar{\delta}\iota^A,\quad\mu=\iota_A\bar{\delta}\iota^A
\end{eqnarray}
with the null tetrad chosen as
\begin{equation}
n^{AA'}=o^A\bar{o}^{A'},\quad l^{AA'}=\iota^A\bar{\iota}^
{A'},\quad m^{AA'}=\iota^A\bar{o}^{A'},\quad \bar{m}^{AA'}=o^A\bar{\iota}^{A'}.
\end{equation}

In what follows, we shall take the null tetrad in our RNdS black hole as
\begin{eqnarray}
&&n^a=(\frac{1}{f},1,0,0),\quad m^a=(0,0,\frac{1}{\sqrt{2}r},-i\frac{\csc\theta}{\sqrt{2}r}),\nonumber\\
&&l^a=(\frac{1}{2},-\frac{f}{2},0,0),\quad \bar{m}^a=(0,0,\frac{1}{\sqrt{2}r},i\frac{\csc\theta}{\sqrt{2}r}),
\end{eqnarray}
whereby the non-vanishing spin-coefficients are given by
\begin{equation}
\gamma=-\frac{f'}{4},\quad \rho=\frac{1}{r},\quad \alpha=-\beta=\frac{\cot\theta}{2\sqrt{2}r},\quad \mu=\frac{f}{2r}.
\end{equation}
By inspection, one can separate the Dirac field as follows
\begin{eqnarray}
&&\phi^1=\frac{1}{\sqrt{2}r}R_-(t,r)_{-\frac{1}{2}}Y_{lm}(\theta,\varphi),\quad \phi^2=\frac{1}{\sqrt{\Box}}R_+(t,r)_{+\frac{1}{2}}Y_{lm}(\theta,\varphi),\nonumber\\
&&\sigma_1=\pm\frac{i}{\sqrt{\Box}}R_+(t,r)_{-\frac{1}{2}}Y_{lm}(\theta,\varphi),\quad \sigma_2=\pm\frac{i}{\sqrt{2}r}R_-(t,r)_{+\frac{1}{2}}Y_{lm}(\theta,\varphi),
\end{eqnarray}
where $_{\pm\frac{1}{2}}Y_{lm}(\theta,\varphi)$ are the spin-$\frac{1}{2}$ weighted spherical harmonics, satisfying
\begin{eqnarray}
(\frac{\partial}{\partial\theta}-i\csc\theta\frac{\partial}{\partial\varphi}+\frac{\cot\theta}{2})_{+\frac{1}{2}}Y_{lm}&=&(l+\frac{1}{2})_{-\frac{1}{2}}Y_{lm},\nonumber\\
(\frac{\partial}{\partial\theta}+i\csc\theta\frac{\partial}{\partial\varphi}+\frac{\cot\theta}{2})_{-\frac{1}{2}}Y_{lm}&=&-(l+\frac{1}{2})_{+\frac{1}{2}}Y_{lm}.
\end{eqnarray}
Whence the corresponding Dirac equation reduces to
\begin{eqnarray}
{[D-iq(A\cdot n)]}R_-+\frac{r}{\sqrt{\Box}}(\frac{l+\frac{1}{2}}{r}\mp i\mathcal{M})R_+&=&0,\nonumber\\
-\frac{\sqrt{\Box}}{2r}(\frac{l+\frac{1}{2}}{r}\pm i\mathcal{M})R_-+{[\Delta-iq(A\cdot l)]}R_+&=&0.
\end{eqnarray}
If we further make the following separation
\begin{equation}
R_\pm=\mathcal{R}_\pm(r)e^{-i\omega t},
\end{equation}
then the above equation can be further expressed as
\begin{eqnarray}\label{D1}
\frac{d\mathcal{R}_\pm}{dr_*}\pm i[\omega-\Phi(r)]\mathcal{R}_\pm+\frac{\sqrt{\Box}}{r}(\frac{l+\frac{1}{2}}{r}\pm i\mathcal{M})\mathcal{R}_\mp&=&0,\\
\frac{d\mathcal{R}_\pm}{dr_*}\pm i[\omega-\Phi(r)]\mathcal{R}_\pm+\frac{\sqrt{\Box}}{r}(\frac{l+\frac{1}{2}}{r}\mp i\mathcal{M})\mathcal{R}_\mp&=&0,\label{D2}
\end{eqnarray}
where the tortoise coordinate is defined as $r_*=\int\frac{dr}{f}$ and the electric potential energy is given by $\Phi(r)=\frac{qQ}{r}$. Whence it is not hard to see there are two sets of independent asymptotic solutions
\begin{eqnarray}
&&\mathcal{R}_+\sim e^{-i[\omega-\phi(r_h)]r_*},\quad \mathcal{R}_-\sim \sqrt{\Box}e^{-i[\omega-\phi(r_h)]r_*};\label{in}\\
&&\mathcal{R}_+\sim \sqrt{\Box}e^{i[\omega-\phi(r_h)]r_*},\quad \mathcal{R}_-\sim e^{i[\omega-\phi(r_h)]r_*}\label{out}
\end{eqnarray}
near any horizon $r_h$.

Now let us focus on the region between the event and cosmological horizons, where $r_*$ can be integrated out explicitly as
\begin{equation}
r_*=-\frac{1}{2\kappa_c}\ln(1-\frac{r}{r_c})+\frac{1}{2\kappa_+}\ln(\frac{r}{r_+}-1)-\frac{1}{2\kappa_-}\ln(\frac{r}{r_-}-1)+\frac{1}{2\kappa_o}\ln(1-\frac{r}{r_o}).
\end{equation}
If we require that the full solution approaches the asymptotic solution (\ref{in}) near the event horizon and (\ref{out}) near the cosmological horizon, respectively, then the equation of motion will lead to a set of discrete frequencies, which is the so-called QNMs.


On the other hand, by performing the gauge transformation with $d\lambda=\frac{Q}{r}dr_*$, and the coordinate transformation to the outgoing coordinates $(u,r)$ with $u$ defined as $u=t-r_*$, we can analytically continue our metric and electric potential across the Cauchy horizon. Furthermore, in terms of the outgoing coordinates, we have
\begin{equation}
n^a=(\frac{\partial}{\partial r})^a,\quad l^a=(\frac{\partial}{\partial u})^a-\frac{1}{2}f(\frac{\partial}{\partial r})^a,
\end{equation}
which implies that our null tetrad as well as the corresponding dyad is also smooth across the Cauchy horizon. However, it is noteworthy that the above quasi-normal solution, when analytically continued into the black hole, generically has both the outgoing mode
\begin{equation}
(\phi^1,\phi^2,\sigma_1,\sigma_2) \sim e^{-i\omega u},
\end{equation}
and the ingoing mode
\begin{equation}
(\phi^1,\sigma_2) \sim e^{-i\omega u}(r-r_-)^{\frac{i[\omega-\Phi(r_-)]}{\kappa_-}+\frac{1}{2}}, \quad (\phi^2,\sigma_1) \sim e^{-i\omega u}(r-r_-)^{\frac{i[\omega-\Phi(r_-)]}{\kappa_-}-\frac{1}{2}}
\end{equation}
near the Cauchy horizon. Obviously, $(\phi^2,\sigma_1)$ is the most dominant part to prevent one from passing through the Cauchy horizon. To be more precise, note that the energy-momentum tensor for our Dirac field can be obtained by the variation of the action with respect to the metric as $T_{ab}=\frac{2}{\sqrt{-g}}\frac{\delta S}{\delta g^{ab}}$, i.e.,
\begin{eqnarray}
T_{AA'BB'}= && -\frac{i}{4}(\phi_A\bar{D}_{BB'}\bar{\phi}_{A'}-\bar{\phi}_{A'}D_{BB'}\phi_A+\phi_B\bar{D}_{AA'}\bar{\phi}_{B'}-\bar{\phi}_{B'}D_{AA'}\phi_B\nonumber\\
&&-\bar{\sigma}_AD_{BB'}\sigma_{A'}+\sigma_{A'}\bar{D}_{BB'}\bar{\sigma}_A-\bar{\sigma}_BD_{AA'}\sigma_{B'}+\sigma_{B'}\bar{D}_{AA'}\bar{\sigma}_B).
\end{eqnarray}
Although it is apparent that the energy-momentum tensor involves the product of our Dirac field with its first derivative, which is different from the case for the Bose fields, where instead the square of the first derivative is involved, we can still follow the similar argument in \cite{DERS} to obtain the exactly same criterion for the violation of SCC. Namely, one can extend this mode across the Cauchy horizon such that SCC is violated if and only if
\begin{equation}
\beta\equiv-\frac{\text{Im}(\omega)}{\kappa_-}> \frac{1}{2}.
\end{equation}
As such, if one can find a quasi-normal mode with $\beta <\frac{1}{2}$, then SCC is respected. So for this purpose, we are only required to focus on the lowest-lying quasi-normal mode. In what follows, we shall focus exclusively on the massless case.

\section{Numerical scheme and relevant results}\label{numerics}
In this section, we shall extract the QNMs by the time domain analysis of the numerical solutions to our Dirac equation in the double null coordinates, and show its implication to SCC.
\subsection{Numerical scheme}
\begin{figure}
  \centering
  \includegraphics[width=6.0cm]{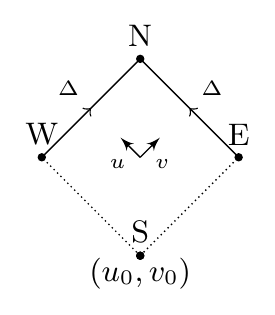}\\
  \caption{Crank-Nicolson method to obtain the data at $N$ from the data at $E$ and $W$ for the Dirac field.}\label{FD}
\end{figure}

In terms of the double null coordinates $(u,v)$ with $v=t+r_*$, we have
\begin{equation}
n^a=\frac{2}{f}(\frac{\partial}{\partial v})^a,\quad l^a=(\frac{\partial}{\partial u})^a.
\end{equation}
Accordingly, the Dirac equation reads
\begin{eqnarray}\label{dnDirac}
(\partial_v-iqA_v)R_-+\frac{\sqrt{\Box}}{2r}(\frac{l+\frac{1}{2}}{r}-i\mathcal{M})R_+&=&0,\nonumber\\
-\frac{\sqrt{\Box}}{2r}(\frac{l+\frac{1}{2}}{r}+i\mathcal{M})R_-+(\partial_u-iqA_u)R_+&=&0.
\end{eqnarray}
Obviously, the numerical solutions to these coupled first order partial differential equations are amenable perfectly to forward Euler method in both $u$ and $v$ directions. But in order to suppress the numerical error and ensure the numerical stability, we would like to solve the above equations by employing Crank-Nicolson method along both $u$ and $v$ directions, whereby a first order differential equation
\begin{equation}
\frac{dy}{dx}=F(y,x)
\end{equation}
will be approximated by
\begin{equation}
\frac{y(x+\triangle)-y(x)}{\triangle}=\frac{1}{2}(F(y(x+\triangle),x+\triangle)+F(y(x),x))+O(\triangle^2).
\end{equation}
Accordingly as illustrated in Fig.\ref{FD}, we can obtain both $R_+$ and $R_-$ at $N$ from the corresponding data at $E$ and $W$ by solving the coupled algebraic equations. In addition, to set off for our numerical evolution, we are required to prescribe the initial value for our Dirac field. For simplicity but without loss of generality, we set the initial value as
\begin{eqnarray}
R_+(0,v)&=&\frac{1}{\sqrt{2\pi}w_1}e^{-\frac{(v-v_c)^2}{2w_1^2}},\nonumber\\
R_+(u,0)&=&\frac{1}{\sqrt{2\pi}w_2}e^{-\frac{(u-u_c)^2}{2w_2^2}}
\end{eqnarray}
with $R_-$ obtained readily by solving our Dirac equation (\ref{dnDirac}) on the initial double null surfaces.

Then the spectrum of low-lying QNMs can be extracted from the equally elapsed late time data $\hat{R}_+(t_p)=R_+(t_0+p\triangle,r_*=0)$ by Prony method\cite{BCGS}. The convergence of our numerics is examined by reducing the finite difference step length $\triangle$. We have also checked our numerics by using the generalized eigenvalue method developed in \cite{Jansen}. Below we shall focus only on the massless Dirac field, although it is obvious that our numerical scheme can also be applied to the massive case.

\subsection{Relevant results}
\begin{table}[htbp]
\centering
\begin{tabular}{cccc}
\toprule
$Q/Q_{\text{max}}$ & $0.991$ & $0.996$ & $0.999$\\ \colrule
$l=\frac{1}{2}$ & \makecell{$(0,-0.347012)$ \\ $(\pm1.32464,-0.494652)$} & \makecell{$(0,-0.574942)$ \\ $(\pm2.20126,-0.813853)$} & \makecell{$(0,-1.26508)$\\$(0,-1.38375)$\\$(\pm4.85209,-1.78313)$} \\
$l=\frac{11}{2}$ & $(\pm8.29423,-0.500452)$ & $(\pm13.8098,-0.822106)$ & $(\pm30.4782,-1.79874)$ \\
$l=\frac{21}{2}$ & $(\pm15.2187,-0.500545)$ & $(\pm25.3398,-0.822236)$ & $(\pm55.9264,-1.79898)$ \\
WKB & $(\quad\quad\quad\quad,-0.500588)$ & $(\quad\quad\quad\quad,-0.822298)$ & $(\quad\quad\quad\quad,-1.79910)$ \\\botrule
\end{tabular}
\caption{The low-lying QNMs $\frac{\omega}{\kappa_-}$ for the case of $q=0$ and $\Lambda=0.005$.}
\label{tab1}
\end{table}

\begin{table}[htbp]
\centering
\begin{tabular}{cccc}
\toprule
$Q/Q_{\text{max}}$ & $0.991$ & $0.996$ & $0.999$\\ \colrule
$l=\frac{1}{2}$ & $(\pm1.27161,-0.455633)$ & $(\pm2.13224,-0.754621)$ & \makecell{$(0,-1.43638)$\\$(\pm4.7395,-1.66474)$} \\
$l=\frac{11}{2}$ & $(\pm7.88119,-0.462177)$ & $(\pm13.2446,-0.764070)$ & $(\pm29.4833,-1.68278)$ \\
$l=\frac{21}{2}$ & $(\pm14.4580,-0.462289)$ & $(\pm24.2982,-0.764230)$ & $(\pm54.0909,-1.68308)$ \\
WKB & $(\quad\quad\quad\quad, -0.462340)$ & $(\quad\quad\quad\quad, -0.764303)$ & $(\quad\quad\quad\quad, -1.68321)$ \\\botrule
\end{tabular}
\caption{The low-lying QNMs $\frac{\omega}{\kappa_-}$ for the case of $q=0$ and $\Lambda=0.06$.}
\label{tab2}
\end{table}

\begin{table}[htbp]
\centering
\begin{tabular}{cccc}
\toprule
$Q/Q_{\text{max}}$ & $0.991$ & $0.996$ & $0.999$\\ \colrule
$l=\frac{1}{2}$ & $(\pm1.08709,-0.359397)$ & $(\pm1.88074,-0.608629)$ & $(\pm4.29788,-1.37227)$ \\
$l=\frac{11}{2}$ & $(\pm6.63585,-0.363011)$ & $(\pm11.5146,-0.613696)$ & $(\pm26.3697,-1.38076)$ \\
$l=\frac{21}{2}$ & $(\pm12.1699,-0.363079)$ & $(\pm21.1186,-0.613790)$ & $(\pm48.3661,-1.38091)$ \\
WKB & $(\quad\quad\quad\quad, -0.363108)$ & $(\quad\quad\quad\quad, -0.613832)$ & $(\quad\quad\quad\quad, -1.38098)$ \\\botrule
\end{tabular}
\caption{The low-lying QNMs $\frac{\omega}{\kappa_-}$ for the case of $q=0$ and $\Lambda=0.14$.}
\label{tab3}
\end{table}

\begin{table}[htbp]
\centering
\begin{tabular}{cccc}
\toprule
$Q/Q_{\text{max}}$ & $0.991$ & $0.996$ & $0.999$\\ \colrule
$l=\frac{1}{2}$ & $(\pm0.512186,-0.147400)$ & $(\pm1.12816,-0.306637)$ & $(\pm2.98040,-0.777079)$ \\
$l=\frac{11}{2}$ & $(\pm3.08308,-0.147759)$ & $(\pm6.81611,-0.30766)$ & $(\pm18.0785,-0.778279)$ \\
$l=\frac{21}{2}$ & $(\pm5.65269,-0.147767)$ & $(\pm12.4980,-0.30768)$ & $(\pm33.1511,-0.778302)$ \\
WKB & $(\quad\quad\quad\quad, -0.147770)$ & $(\quad\quad\quad\quad, -0.307689)$ & $(\quad\quad\quad\quad, -0.778313)$ \\\botrule
\end{tabular}
\caption{The low-lying QNMs $\frac{\omega}{\kappa_-}$ for the case of $q=0$ and $\Lambda=0.20$.}
\label{tab4}
\end{table}

In what follows, we shall work with the units in which $M=1$ with $Q_\text{max}$ corresponding to the charge of the extremal black hole, where $r_+$ coincides with $r_-$. In addition, due to the limited computational resources, we shall restrict ourselves onto some sample points in the near-extremal regime, which also suffices for our purpose. We would like to first present the low-lying QNMs for the neutral Dirac field in Table.\ref{tab1},\ref{tab2},\ref{tab3}, and \ref{tab4}. Here we list the low-lying QNMs for each $l$ until the appearance of the first pair of complex photon sphere modes. In addition, we only show the imaginary part for the large $l$ WKB limit, since the corresponding real part does not converge (See Appendix A). As expected, the resulting spectrum of QNMs is symmetric with respect to the imaginary axis of the $\omega$-plane. On the other hand, for each sample case, $l=\frac{1}{2}$ mode always dominates, no matter whether it serves as a purely imaginary mode or a pair of complex photon sphere modes. More importantly, for a fixed $\Lambda$, when one cranks up the charge of our black hole towards the extremal one, SCC is eventually violated.

\begin{figure}
  \centering
  \includegraphics[width=12.0cm]{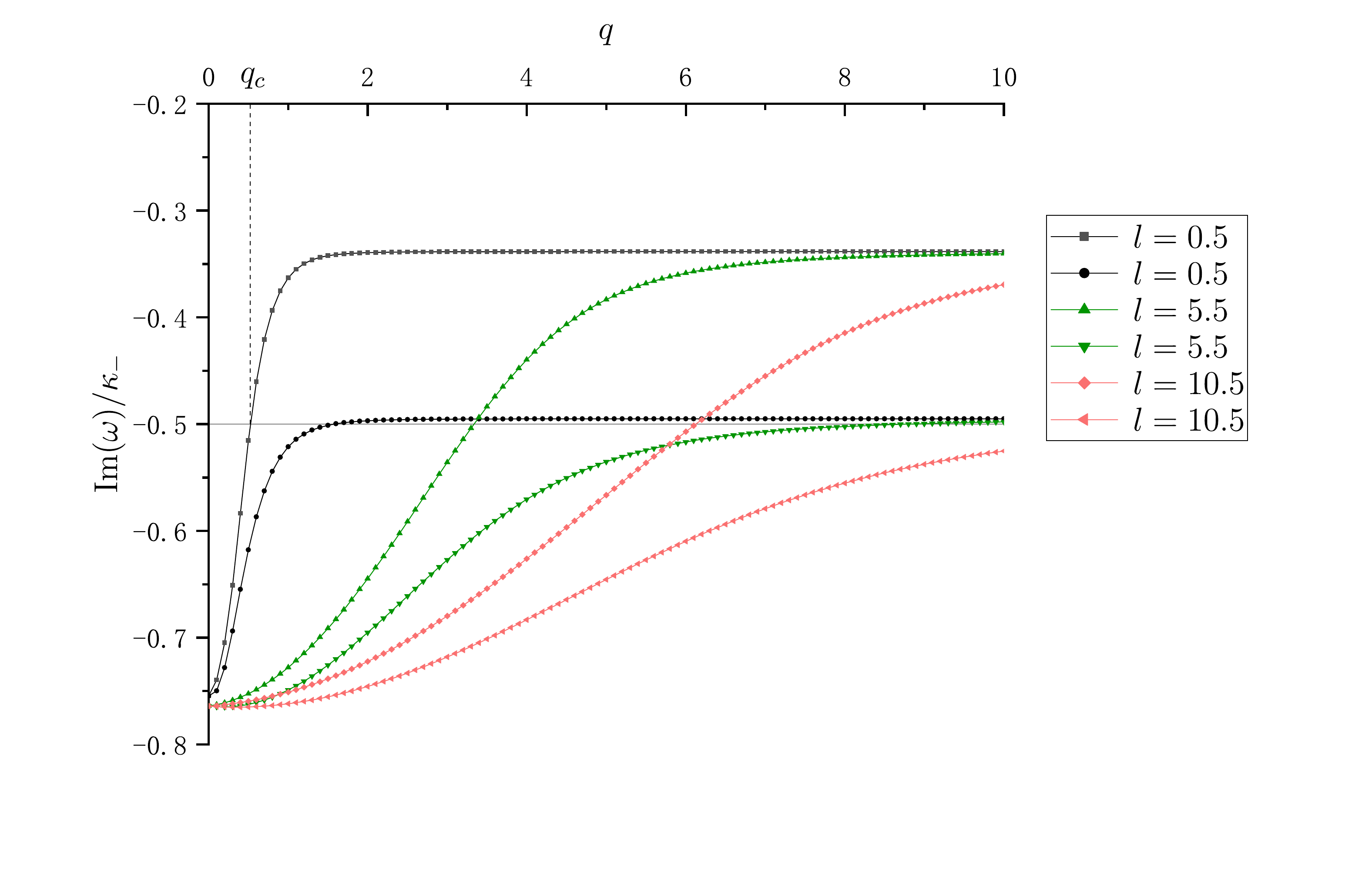}\\
  \caption{The low-lying QNMs for $\Lambda=0.06$ and $Q/Q_\text{max}=0.996$.}\label{small}
\end{figure}

\begin{figure}
  \centering
  \includegraphics[width=12.0cm]{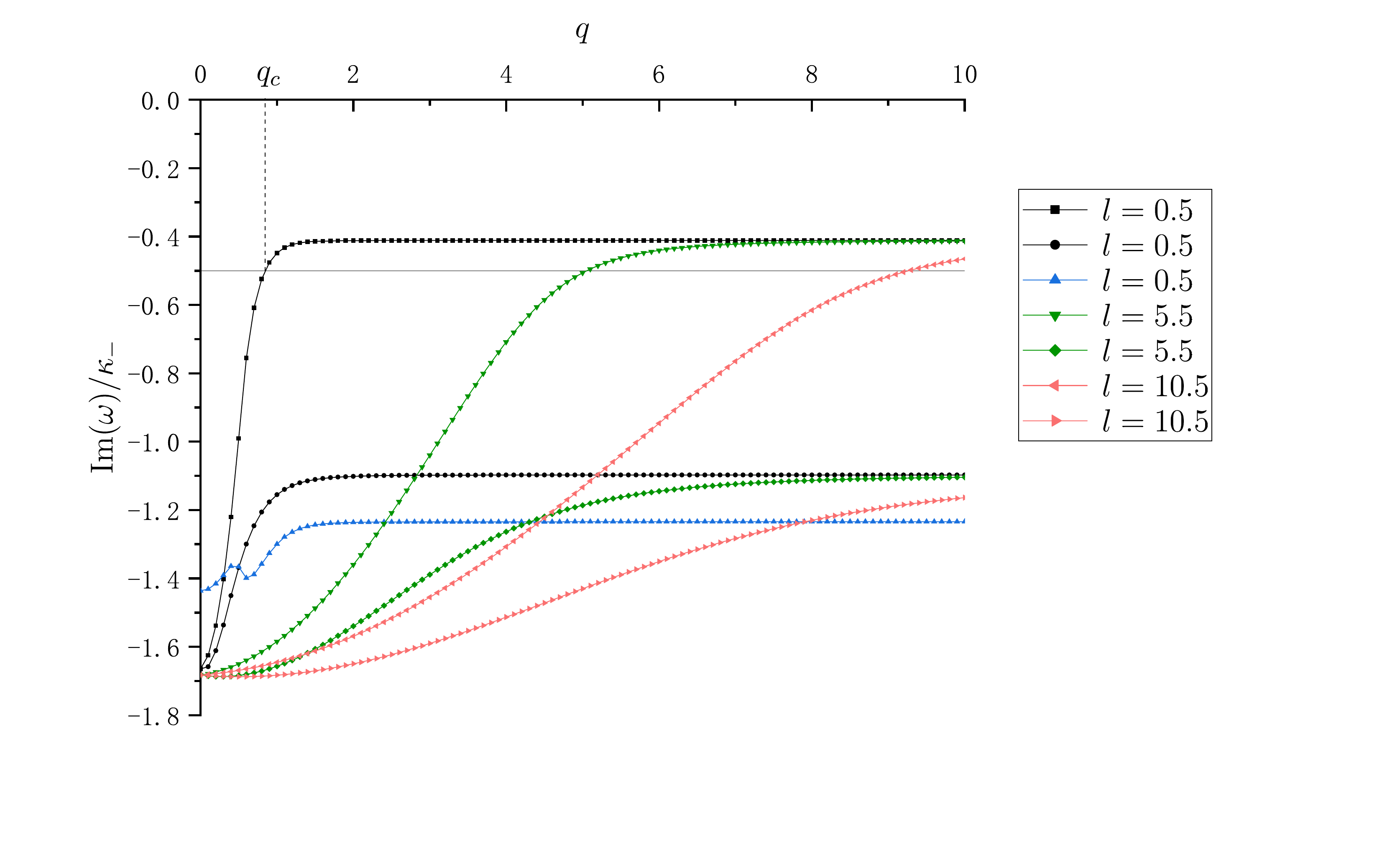}\\
  \caption{The low-lying QNMs for $\Lambda=0.06$ and $Q/Q_\text{max}=0.999$, where the blue line denotes the variation of the initially purely imaginary mode with the charge of our Dirac field.}\label{large}
\end{figure}

Next we shall present how the spectrum of low-lying QNMs varies with the charge of our Dirac field. As a demonstration, we only plot the relevant results of $\Lambda=0.06$ in Fig.\ref{small} and Fig.\ref{large} for $Q/Q_\text{max}=0.996$ and $Q/Q_\text{max}=0.999$, respectively. As shown in Fig.\ref{small}, once we charge our Dirac field, each pair of photon sphere modes lose the left and right symmetry with respect to the imaginary axis of the $\omega$-plane. In particular, the large $q$ behavior of the photon sphere modes turns out to be in good agreement with our analytic result, namely one mode goes into the black hole family with the imaginary part of approaching $-\frac{\kappa_+}{2}$ and the other mode goes into the cosmological family with the imaginary part approaching $-\frac{\kappa_c}{2}$ (See Appendix B). More relevant to our purpose, we find that the SCC is recovered by the $l=\frac{1}{2}$ black hole family mode when the charge of our Dirac field is greater than a critical charge $q_c\approx 0.530$. As illustrated in Fig.\ref{large}, such an observation also applies to the case for the more near-extremal black hole except that the critical charge is increased to $q_c\approx 0.845$.

\section{Conclusion and discussion}\label{end}
To see whether the Dirac field can save SCC out of the RNdS black hole, we first derive from scratch the criterion for the dominant QNMs of our Dirac field to violate SCC around the RNdS black hole. As a result, we find that such a criterion is exactly the same as that for other Bose fields, although the energy-momentum tensor of our Dirac field demonstrates qualitatively different structure from those for other Bose fields. Then to extract the QNMs by prony method, we develop our numerical scheme by Crank-Nicolson method, which turns out to be naturally suitable to the temporal evolution of our Dirac field in the double null coordinates. In particular, our numerical result shows that for a fixed RNdS black hole, SCC can be recovered by the $l=\frac{1}{2}$ black hole family mode when the charge of our Dirac field is greater than some critical value, which is also consistent with the WKB result for the large $q$ limit. In addition, the afore-mentioned critical value is increased as one approaches the extremal RNdS black hole.

\begin{acknowledgments}
We are grateful to Yuyu Mo and Jieru Shan for their helpful discussions in the early stage of this project.
This work is partially supported by NSFC with Grant No.11475179, No.11675015, No.11775022, and No. 11875095 as
well as by FWO-Vlaanderen through the project G020714N, G044016N, and G006918N.  BW acknowledges the support by NSFC with Grant
No.11575109. HZ is an individual FWO Fellow supported by 12G3515N and by the Vrije Universiteit Brussel through the Strategic Research Program ``High-Energy Physics".
\end{acknowledgments}

\section*{Appendix A: WKB approximation for the large $l$ limit}\label{A}
As to the massless Dirac field, Eq.(\ref{D1}) is actually equivalent to Eq.(\ref{D2}). In particular, to make it amenable to the WKB treatment\cite{IW,Iyer,Konoplya1,Konoplya2}, we like to follow \cite{Chandra} to rewrite it as
\begin{equation}\label{susy}
[\frac{d}{d\hat{r}_*}\pm W(r)]\mathcal{Z}_\pm=-i\omega \mathcal{Z}_\mp
\end{equation}
with $\mathcal{Z}_\pm=\mathcal{R}_+\pm \mathcal{R}_-$, $\frac{d\hat{r}_*}{dr_*}=1-\frac{\Phi}{\omega}$, and $W(r)=(1-\frac{\Phi}{\omega})^{-1}\frac{\sqrt{\Box}}{r^2}(l+\frac{1}{2})$. Whence we can further obtain a pair of decoupled equations as
\begin{equation}
(\frac{d^2}{d\hat{r}_*^2}+\omega^2)\mathcal{Z}_\pm=V_\pm(r)\mathcal{Z}_\pm
\end{equation}
with $V_\pm(r)=W^2\mp\frac{dW}{d\hat{r}_*}$. Obviously this pair of equations give rise to the same spectrum of QNMs because $\mathcal{Z}_\pm$ are related to each other as a supersymmetric partnership through (\ref{susy}). With this preparation, we can apply the third-order WKB approximation formula to the above equation with $V_+$ to extract the corresponding low-lying QNMs, which is believed to be highly accurate in the large $l$ limit\cite{Iyer}.

\section*{Appendix B: WKB approximation for the large $q$ limit}\label{B}
To obtain the relevant result in the large $q$ limit, we prefer to work with the ingoing coordinates $(v,r)$ with $v$ defined as $v=t+r_*$. At the same time, we make a gauge transformation such that $A_a=-\frac{Q}{r}(dv)_a$. With the assumption $R_\pm=e^{-i\omega v}\mathcal{R}_\pm(r)$, the massless Dirac equation reads
\begin{eqnarray}
f\frac{d\mathcal{R}_+}{dr}+\frac{\sqrt{f}}{r}(l+\frac{1}{2})\mathcal{R}_-&=&0,\label{minus}\\
f\frac{d\mathcal{R}_-}{dr}-2i[\omega-\Phi(r)]\mathcal{R}_-+\frac{\sqrt{f}}{r}(l+\frac{1}{2})\mathcal{R}_+&=&0.\label{plus}
\end{eqnarray}
Obtaining the expression of $\mathcal{R}_-$ in terms of $\mathcal{R}_+$ by Eq.(\ref{minus}) and plugging it into Eq.(\ref{plus}), we end up with the decoupled equation for $\mathcal{R}_+$ as\footnote{This is also the equation we use in the generalized eigenvalue method to extract the spectrum of low-lying QNMs.}
\begin{equation}
    -4r^2f\mathcal{R}_+''+2ir[4r(\omega-\Phi)+2if+irf']\mathcal{R}_+'+(2l+1)^2\mathcal{R}_+=0,\label{largeq}
\end{equation}
where the prime denotes the derivative with respect to $r$. Now we like to follow \cite{DRS2} by postulating the expansion in $\frac{1}{q}$ as\footnote{For our purpose, here we ignore the potential non-perturbative terms if any.}
\begin{eqnarray}
\mathcal{R}_+&=&(1-\frac{r}{r_c})^{\frac{1}{2}-i\frac{[\omega-\Phi(r_c)]}{\kappa_c}}e^{-q \psi(r)}\sum_{n=0}^{+\infty}\frac{\mathcal{R}_+^{n}(r)}{q^n},\\
\omega&=&\sum_{n=-1}^{+\infty}\frac{\omega^{(n)}}{q^n},
\end{eqnarray}
which can be solved order by order in $\frac{1}{q}$ by substituting the above ansatz to Eq.(\ref{largeq}).
To leading order, we find two possibilities
\begin{eqnarray}
\omega^{(-1)}_{+}&=&\frac{Q}{r_+},\\
\omega^{(-1)}_{c}&=&\frac{Q}{r_c},
\end{eqnarray}
which are dubbed as the black hole family and cosmological family, respectively.
Accordingly, the corresponding equation for $\psi$ is given by
\begin{eqnarray}
\psi_+'&=&\frac{i Q [2\kappa_cr_c(r-r_c)(r-r_+)+r(r_c-r_+)f]}{\kappa_cr_cr_+r(r_c-r)f},\\
\psi_c'&=&0.
\end{eqnarray}
Because we care only about the large $q$ limit, here we just write the first few corrections to $\omega$ as
\begin{equation}
\omega_+^{(0)}=-\frac{i\kappa_+}{2}, \quad \omega_+^{(1)}=\frac{(1+2 l)^2\kappa_+}{8 Q},\quad \omega_+^{(2)}=0,
\end{equation}
and
\begin{equation}
\omega_c^{(0)}=-\frac{i\kappa_c}{2},\quad \omega_c^{(1)}=-\frac{(1+2 l)^2\kappa_c}{8 Q}, \quad \omega_c^{(2)}=0.
\end{equation}

\end{document}